\documentclass[aps,reprint,twocolumn,superscriptaddress,10pt]{revtex4-1}

\usepackage{amsfonts}
\usepackage{amsmath}
\usepackage{amssymb}
\usepackage{gensymb}
\usepackage{graphicx}
\usepackage{dcolumn}
\usepackage{mciteplus}
\usepackage{euscript}
\usepackage{multirow}
\usepackage{color}
\usepackage{siunitx}
\providecommand{\U}[1]{\protect\rule{.1in}{.1in}}
\newcommand{\redtx}[1]{\textcolor{red}{}}
\newcommand{\cecof}{CeCo$_{5}$}

\newcommand{\ceco}{Ce$_{2}$Co$_{17}$}
\newcommand{\rceco}{$R$-Ce$_{2}$Co$_{17}$}
\newcommand{\hceco}{$H$-Ce$_{2}$Co$_{17}$}

\newcommand{\kso}{$K_\text{so}$}

\newcommand{\morb}{$M_{l}$}

\newcommand\etal{$\textit{et al.}$}
\def\abinitio{$\emph{ab initio}$}

\def\mspin{$m_{s}$}

\def\morb{$m_{l}$}
\newcommand{\mevfu}{\meV\per{f.u.}}
\newcommand{\mubfu}{\mu_B\per{f.u.}}
\newcommand{\mevat}{\meV\per{atom}}
\newcommand{\mjmc}{\,MJ$m^{-3}$}

\def\mub{{$\mu_B$}}

\def\tc{T_\text{C}}

\newcommand{\req}[1]{Eq.~(\ref{#1})}

\newcommand{\rfig}[1]{Fig.~\ref{#1}}
\newcommand{\rtbl}[1]{Table~\ref{#1}}

\begin{document}

\title{ Origin of magnetic anisotropy in doped Ce$_2$Co$_{17}$ alloys}

\begin{abstract}
Magnetocrystalline anisotropy (MCA) in doped {\ceco} and other
competing structures was investigated using density functional theory.
We confirmed that the MCA contribution from dumbbell Co sites is very
negative.  Replacing Co dumbbell atoms with a pair of Fe or Mn atoms
greatly enhance the uniaxial anisotropy, which agrees quantitatively
with experiment, and this enhancement arises from electronic-structure
features near the Fermi level, mostly associated with dumbbell sites.
With Co dumbbell atoms replaced by other elements, the variation of
anisotropy is generally a collective effect and contributions from
other sublattices may change significantly.  Moreover, we found that
Zr doping promotes the formation of 1-5 structure that exhibits a
large uniaxial anisotropy, such that Zr is the most effective element
to enhance MCA in this system.
\end{abstract}

\eid{identifier}
\date{\today}
\author{Liqin Ke}
\email[Corresponding author: ]{liqinke@ameslab.gov}
\affiliation{Ames Laboratory, U.S. Department of Energy, Ames, Iowa 50011, USA}
\author{D. A. Kukusta}
\affiliation{Ames Laboratory, U.S. Department of Energy, Ames, Iowa 50011, USA}
\affiliation{Institute for Metal Physics, 36 Vernadsky Street, 03142 Kiev, Ukraine}
\author{Duane D. Johnson}
\affiliation{Ames Laboratory, U.S. Department of Energy, Ames, Iowa 50011, USA}
\affiliation{Departments of Materials Science $\&$ Engineering and Physics, Iowa State University, Ames, Iowa 50011-2300}

\maketitle
\section{Introduction}

The quest for novel high energy permanent magnet without critical
elements continues to generate great
interest~\cite{mccallum.arms2014}.  While a rare-earth-free permanent
magnet is appealing, developing a Ce-based permanent magnet is also
very attractive, because among rare-earth elements Ce is most abundant
and relatively cheap. Among Ce-Co systems, Ce$_{2}$Co$_{17}$ has
always attracted much attention due to its large Curie temperature
$\tc$ and magnetization $M$. The weak point of {\ceco} is its rather
small easy-axis magnetocrystalline anisotropy (MCA), which must be
improved to use as an applicable permanent magnet.

\begin{figure}[ht]
\begin{tabular}{c}
  \includegraphics[width=.45\textwidth,clip,angle=0]{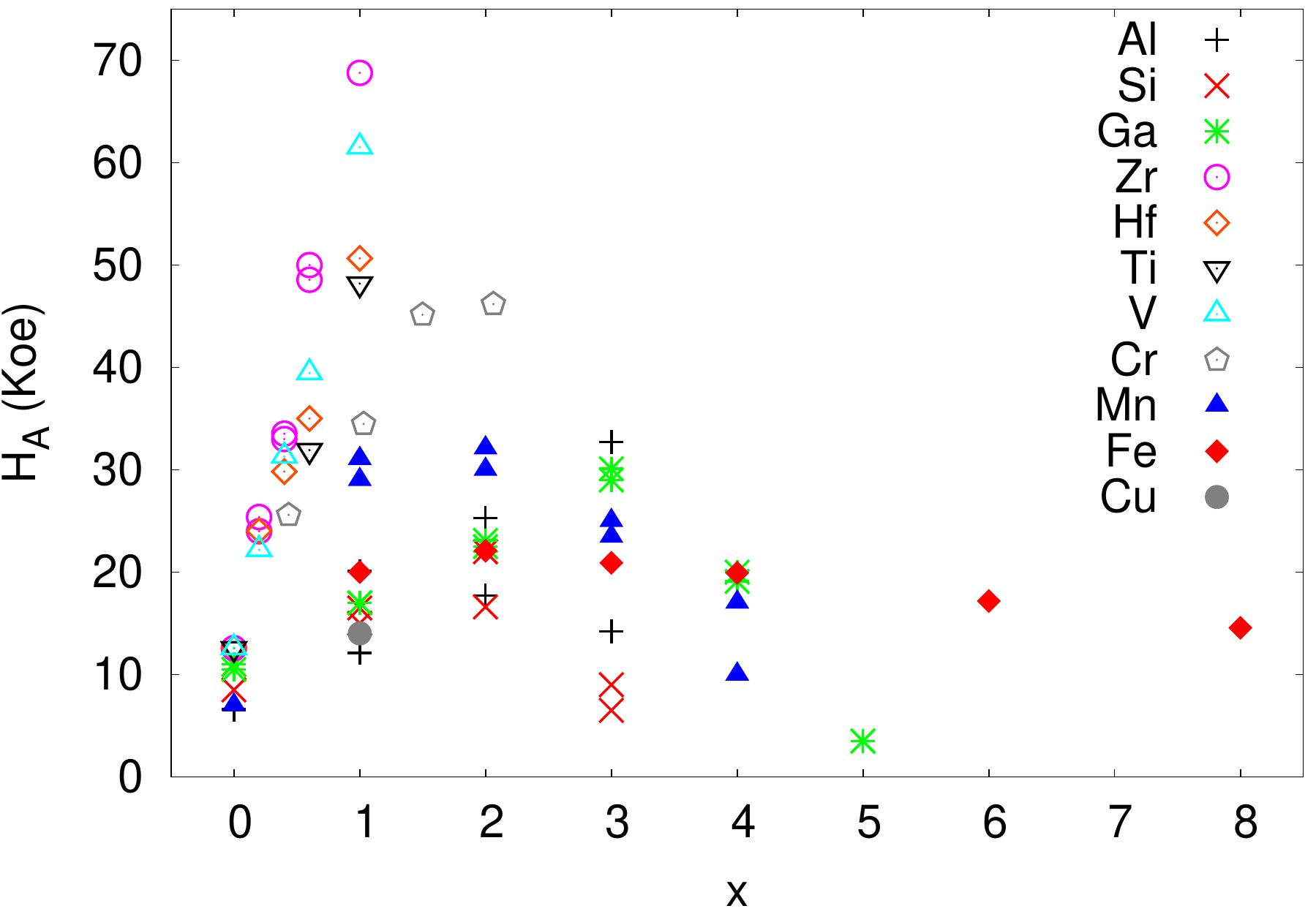} \\
\end{tabular}%
\caption{Experimental anisotropy fields $H_\text{A}$ in
  Ce$_2T_x$Co$_{17-x}$ with $T$=Al~\cite{hu.jac1999,shen.jap1999},
  Si~\cite{hu.jac1999,wei.pb1999}, Ga~\cite{hu.jac1999,wei.jac1998},
  Zr~\cite{fujii.ssc1982}, Hf~\cite{fujii.ssc1982},
  V~\cite{fujii.jap1982}, Cr~\cite{fujii.jap1982},
  Mn~\cite{sun.jpcm2000,fujii.jap1982}, Fe~\cite{fujii.jap1982}, and
  Cu~\cite{fujii.jap1982}. }
\label{fig:expt}
\end{figure}

The anisotropy in {\ceco}, in fact, can be improved significantly
through doping with various elements. Experimental anisotropy field
$H_\text{A}$ measurements by dopant and stoichiometry are shown in
\rfig{fig:expt}. This anisotropy enhancement has been attributed to
the preferential substitution effects of doping
atoms~\cite{buschow.rpp1977,hmm.v04c2}: (i) The four non-equivalent Co
sites contribute differently~\cite{yajima.jpsj1972} to the magnetic
anisotropy in Ce$_{2}$Co$_{17}$. Two out of the 17 Co atoms occupy the
so-called dumbbell sites and have a very negative contribution to
uniaxial anisotropy, leading to the small overall uniaxial anisotropy;
(ii) Doping atoms preferentially replace the dumbbell sites first,
eliminating their negative contribution and increasing the overall
uniaxial anisotropy. The above explanation is supported by the
observation that with many different dopants, the anisotropy field in
Ce$_{2}T_{x}$Co$_{17-x}$ shows a maximum around $x=2$. This
corresponds to the number of dumbbell sites in one formula
unit~\cite{fujii.jap1982}.

Numerous experimental efforts have explored the preferential
substitution effect and site-resolved anisotropy.
Streever~\cite{streever.prb1979} studied the site contribution to the
MCA in {\ceco} using nuclear magnetic resonance and concluded that the
dumbbell sites in {\ceco} have a very negative contribution to
uniaxial anisotropy.  Neutron scattering or M\"ossbauer studies have
suggested that
Fe~\cite{deportes.jlcm1976,perkins.ssc1976,gubbens.pssa1976,streever.prb1979},
Mn~\cite{kuchin.jac2000}, and
Al~\cite{inomata.prb1981,deGroot.jac1996,shen.jap1999} atoms prefer to
substitute at dumbbell sites.

However, it is not clear whether only the preferential substitution
effect plays a role in $H_\text{A}$ enhancement for all doping
elements. For elements such as Zr, Ti, and Hf, the substitution
preference is not well understood. Replacing the dumbbell Co atoms
with a pair of large atoms may not always be the only energetically
favorable configuration. For Mn and Fe, known to substitute at
dumbbell sites, the elimination of negative contributions at those
sites may explain the increase of magnetocrystalline anisotropy
energy~(MAE). It is yet unclear why different elements give a
different amplitude of MAE enhancement or what mechanism provides this
enhancement. For permanent magnet application, Fe and Mn are
particularly interesting because they improve the anisotropy while
preserving the magnetization with $x<2$. Other dopants quickly reduce
the magnetization and Curie temperature. Further tuning of magnetic
properties for compounds based on Fe-or-Mn-doped {\ceco} would benefit
from this understanding.

In this work, we use density functional theory (DFT) to investigate
the origin of the MAE enhancement in doped {\ceco}. By evaluating the
on-site spin-orbit coupling (SOC)
energy~\cite{antropov.ssc2014,ke.prb2015}, we resolved anisotropy into
contributions from atomic sites, spins, and orbital
pairs. Furthermore, we explained the electronic-structure origin of
MAE enhancement.

\section{Calculation details}

\begin{figure*}[tbp]
\includegraphics[width=0.84\textwidth,clip,angle=0]{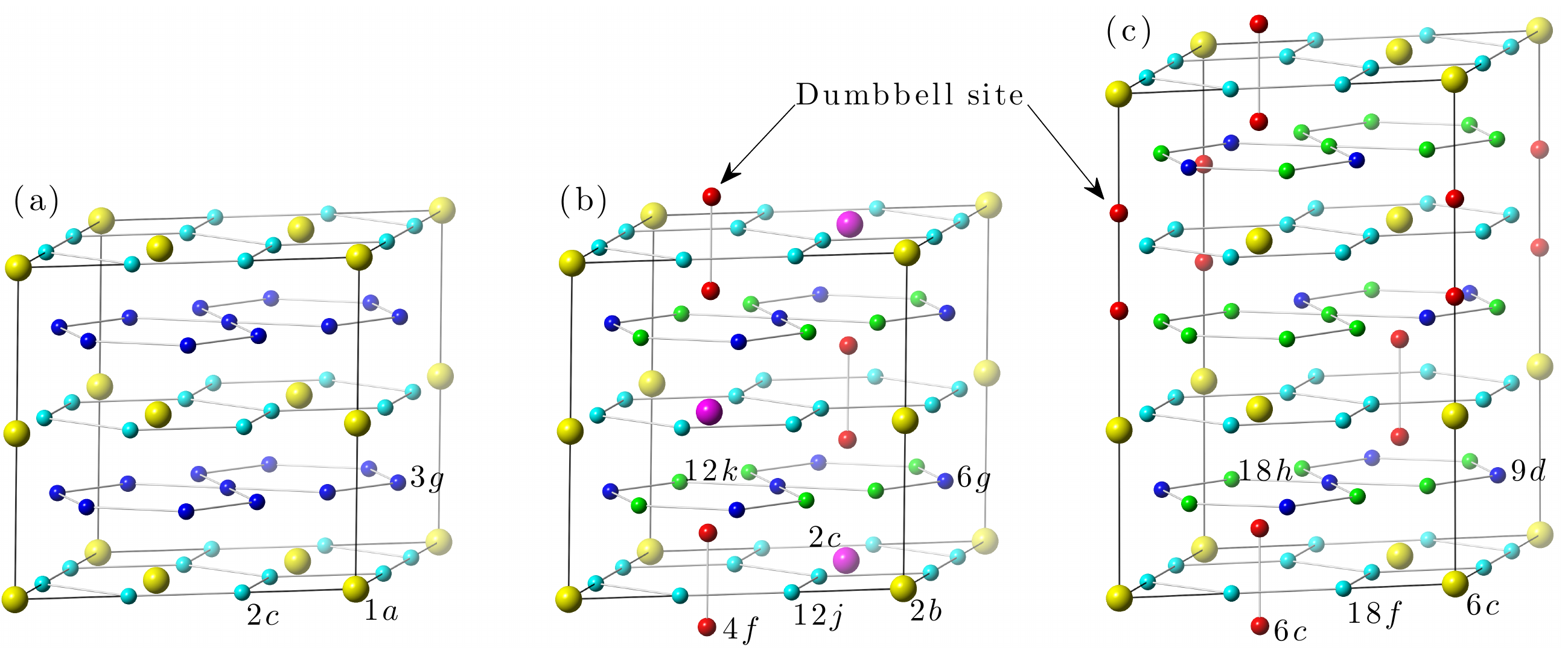}
\caption{ Schematic crystal structures of (a) CeCo$_{5}$, (b)
  hexagonal {\hceco}, and (c) rhombohedral {\rceco}. Ce atoms are
  indicated with large (yellow or magenta colored) spheres. Co atoms
  are denoted by Wyckoff sites. Dumbbell (red) sites are denoted in
  {\hceco} ($4f$ sites) and in {\rceco} ($6c$ sites), and indicated
  further by arrows and label.  We use larger cells for CeCo$_5$ and
  {\rceco} to compare with {\hceco}. }
\label{fig:xtal}
\end{figure*}

\subsection{Crystal structure}

{Ce$_{2}$Co$_{17}$} crystallizes in the hexagonal
Th$_{2}$Ni$_{17}$-type ($P63/mmc$, space group no. 194) structure or
the rhombohedral Zn$_{17}$Th$_{2}$-type ($R\overline{3}mh$, space
group no. 166) structure, depending on growth condition and
doping~\cite{fujii.ssc1982}. As shown in \rfig{fig:xtal}, both
$2$-$17$ structures can be derived from the hexagonal CaCu$_5$-type
($P6/mmm$ space group 191) structure with every third Ce atom being
replaced by a pair of Co atoms (referred to as dumbbell sites). The
two 2-17 structures differ only in the spatial ordering of the
replacement sites.  In the {\cecof} cell, a Ce atom occupies the $1a
(6/mmm)$ site and two Co atoms occupy the $2c (-6m2)$ site, together
forming a Ce-Co basal plane. Three Co atoms occupy the $3g (mmm)$
sites and form a pure Co basal plane.  The primitive cell of hexagonal
{\ceco}(\hceco) contains two formula units while the rhombohedral
{\ceco}(\rceco) contains one. The Co atoms are divided into four
sublattices, denoted by Wyckoff sites $18h$, $18f$, $9d$, and $6c$ in
the rhombohedral structure, and $12k$, $12j$, $6g$, and $4f$ in the
hexagonal structure. The $6c$ and $4f$ sites are the dumbbell sites.
In the $R$-structure, Ce atoms form -Ce-Ce-Co-Co- chains with Co atoms
along the $z$ axis. The $H$-structure has two inequivalent Ce sites,
denoted as $2c$ and $2b$, respectively. Along the $z$ direction,
Ce$_{2b}$ form pure -Ce- atoms chains and Ce$_{2c}$ form
-Ce$_{2c}$-Co-Co- chains with Co dumbbell sites.

\subsection{Computational methods}
We carried out first principles DFT calculations using the Vienna
{\abinitio} simulation package
(VASP)~\cite{kresse.prb1993,kresse.prb1996} and a variant of the
full-potential linear muffin-tin orbital (LMTO)
method~\cite{methfessel.chap2000}. We fully relaxed the atomic
positions and lattice parameters, while preserving the symmetry using
VASP. The nuclei and core electrons were described by the projector
augmented-wave potential~\cite{kresse.prb1999} and the wave functions
of valence electrons were expanded in a plane-wave basis set with a
cutoff energy of \SI{520}{\eV}.  The generalized gradient
approximation of Perdew, Burke, and Ernzerhof was used for the
correlation and exchange potentials.

The MAE is calculated below as $K$=$E_{100}{-}E_{001}$, where
$E_{001}$ and $E_{100}$ are the total energies for the magnetization
oriented along the $[001]$ and $[100]$ directions,
respectively. Positive (negative) $K$ corresponds to uniaxial (planar)
anisotropy.  The spin-orbit coupling is included using the
second-variation procedure~\cite{koelling.jpc1977,shick.prb1997}. The
$k$-point integration was performed using a modified tetrahedron
method with Bl\"ochl corrections.  To ensure the convergence of the
calculated MAE, dense $k$ meshes were used. For example, we used a
$16^3$ $k$-point mesh for the calculation of MAE in {\rceco}.  We also
calculated the MAE by carrying out all-electron calculations using the
full-potential LMTO (FP-LMTO) method to check anisotropy results. To
decompose the MAE, we evaluate the anisotropy of the scaled on-site
SOC energy {\kso}=$\frac{1}{2}\langle V_\text{so}
\rangle_{100}{-}\frac{1}{2}\langle V_\text{so}
\rangle_{001}$. According to second-order perturbation
theory~\cite{antropov.ssc2014,ke.prb2015},
$K\approx\sum_{i}K_\text{so}(i)$, where $i$ indicates the atomic
sites. Unlike $K$, which is calculated from the total energy
difference, {\kso} is localized and can be decomposed into sites,
spins, and subband pairs~\cite{antropov.ssc2014,ke.prb2015}.

\section{Results and discussion}

\subsection{{\ceco}}

\begin{table}[ht]
\caption{Atomic spin {\mspin} and orbital {\morb} magnetic moments
  (\mub/atom) in CeCo$_5$, {\rceco} and {\hceco}. Atomic sites are
  grouped to reflect how the 2-17 structure arises from the 1-5
  structure. Calculated interstitial spin moments are around
  \SI{-1.1}{\mubfu} in {\ceco} and \SI{-0.4}{\mubfu} in
     {\cecof}. Measured magnetization is \SI{26.5}{\mubfu} in {\hceco}
     at 5$K$~\cite{hu.jac1999}, and \SI{7.12}{\mubfu} in
     {\cecof}~\cite{bartashevich.jmmm1996}.  Dumbbell sites are
     denoted as $6c$ and $4f$ in {\rceco} and {\hceco}, respectively.}
\label{tbl:msite}%
\bgroup
\def\arraystretch{1.2}
\begin{tabular}{cc|c|cccc|cccc|c}
  \hline \hline
\\[-1.1em]
{CeCo$_5$}               &    &  $2c$     &    &  \multicolumn{2}{c}{${3g}$}    &     & \multicolumn{3}{c}{$1a$ (Ce)}                  &    &  Total      \\
\mspin                   &    &  1.33     &    &  \multicolumn{2}{c}{1.44}      &     & \multicolumn{3}{c}{-0.76}                      &    &  6.22       \\  
\morb                    &    &  0.14     &    &  \multicolumn{2}{c}{0.12}      &     & \multicolumn{3}{c}{0.30}                       &    &  0.92       \\  
\hline
\\[-1.1em]                                                                                                                               
{\rceco}                 &    &  $18f$    &    &  $18h$       &   $9d$          &     &  $6c$       & \multicolumn{2}{c}{$6c$(Ce)}     &    &  Total       \\
\mspin                   &    &  1.53     &    & 1.43         & 1.52            &     &  1.65       & \multicolumn{2}{c}{ -0.85}       &    &  23.94       \\ 
\morb                    &    &  0.10     &    & 0.09         & 0.07            &     &  0.07       & \multicolumn{2}{c}{  0.35}       &    &   2.17       \\ 
\hline
\\[-1.1em]                                                                                                                               
{\hceco}                 &    &  $12j$    &    & $12k$        & $6g$            &     & $4f$        & $2c$(Ce)     &  $2b$(Ce)         &    &  Total       \\
\mspin                   &    &  1.56     &    & 1.51         & 1.51            &     & 1.65        & -0.84        &  -0.90            &    & 24.50        \\  
\morb                    &    &  0.11     &    & 0.10         & 0.08            &     & 0.07        &  0.38        &   0.42            &    &  2.43        \\  
\\[-1.0em]
\hline\hline
\end{tabular}%
\egroup
\end{table}

Atomic spin and orbital magnetic moments in {\ceco} and {\cecof} are
summarized in \rtbl{tbl:msite}. The calculated magnetization are 25.2
and \SI{25.8}{\mubfu} in {\rceco} and \hceco, respectively, and
\SI{6.75}{\mubfu} in {\cecof}, which agree well with
experiments~\cite{hu.jac1999}. Ce spin couples antiferromagneticlly
with the Co spin. The orbital magnetic moment of Ce is antiparallel to
its spin, which reflects the Hunds' third rule.  In the Ce-Co plane of
{\ceco} the Ce atoms are partially replaced by dumbbell Co atoms and
this leads to an increased moment for the Co atoms (in that plane) as
compared to {\cecof}, The dumbbell sites have the largest magnetic
moment due to its relatively large volume. Calculation shows {\ceco}
has a small uniaxial anisotropy, \SI{0.13}{\mevfu}~(0.09{\mjmc}) and
\SI{0.47}{\mevfu}~(0.30{\mjmc}) for {\rceco} and {\hceco},
respectively. The experimental values fall slightly above the
calculated ones, see \rfig{fig:mae_t2_all}.

To understand the low uniaxial anisotropy in {\ceco}, we resolve the
anisotropy into atomic sites by evaluating{} \kso. The anisotropy
contributions in {\ceco} can be divided into three groups: the pure Co
plane ($3g$ in CeCo$_{5}$, $12k+6g$ in {\hceco}, or $18h+9d$ in
{\rceco}), the Ce-Co plane, and the Co dumbbell pairs. We found that
the MAE contributions from these three groups in the two 2-17
structures are very similar: the dumbbell Co sites have a very
negative contribution to uniaxial anisotropy; the pure-Co basal plane
has a negligible or even slightly negative contribution to the
uniaxial anisotropy; only the Ce-Co basal plane provides uniaxial
anisotropy in{} \ceco. The two inequivalent Ce sites contribute
differently to the uniaxial anisotropy in $H$-Ce$_{2}$Co$_{17}$
structure. Ce($2b$) supports uniaxial anisotropy while Ce($2c$) moment
prefer to be in-plane. However, the total contribution from the two Ce
sites is positive, as in the $R$-structure.

Intrinsic magnetic properties and the effect of doping on them are
very similar in the two 2-17 structures. We only discuss the results
calculated using the $R$-structure because it has a smaller primitive
cell than the $H$-structure, and the most interesting substituents, Fe
and Mn, promote its formation~\cite{fujii.jap1982}.

\subsection{ MAE in Ce$_2$$T_2$Co$_{15}$}

\begin{figure}[ht]
\begin{tabular}{c}
 \includegraphics[width=.42\textwidth,clip,angle=0]{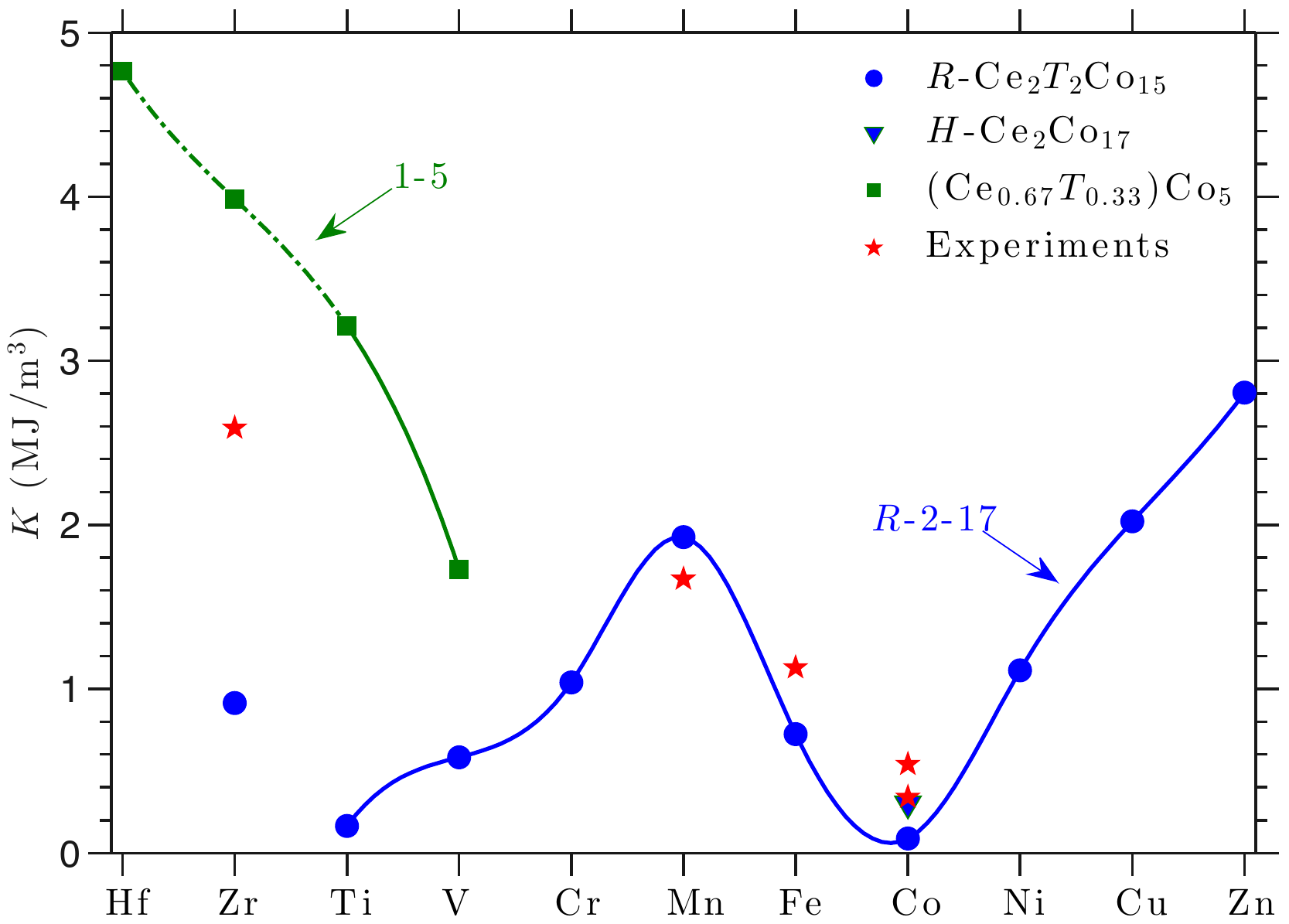}
\end{tabular}%
\caption{ Magnetic anisotropy in Ce$_2T_2$Co$_{15}$ and
  Ce$_{0.66}T_{0.33}$Co$_5$ with $T$=Ti, V, Cr, Mn, Fe, Co, Ni, Cu,
  Zr, and Hf. In Ce$_2T_2$Co$_{15}$, $T$ atoms occupy the dumbbell
  sublattice. The Ce$_{0.66}T_{0.33}$Co$_5$ structure was obtained by
  replacing the pair of dumbbell Co atoms in the original {\ceco} with
  a single $T$ atom. $K$ values derived from experimental $H_\text{A}$
  measurements~\cite{fujii.jap1982,sun.jpcm2000} by using
  $K$=$\frac{1}{2}\mu_0M_\text{s}H_\text{A}$ are also shown.}
\label{fig:mae_t2_all}
\end{figure}

We first calculate the MAE in Ce$_2T_2$Co$_{15}$ with a variety of
doping elements $T$, by assuming the pair of Co dumbbell atoms is
replaced by a pair of doping atoms. The calculated MAE as a function
of doping elements for $T$=Zr and $3d$ elements is shown in
\rfig{fig:mae_t2_all}. Fe and Mn doping increase the MAE, aligning
with with experimental results. However, the MAE calculated for light
$d$ elements $T$=Ti, V, and Zr are rather small while experiments show
that large enhancements of MAE can be achieved with a small amount of
doping of those elements. Interestingly, large MAE values are obtained
in Ce$_2T_2$Co$_{15}$ with $T$=Cu or Zn. In fact, a small amount of Cu
are often added to the alloy to improve the coercivity and the
enhancement had been interpreted as precipitation hardening by Cu. It
may not be unexpected that the enhancement of coercivity may also
partially arise from the increase of MAE, although Cu atoms had been
reported to randomly occupy all Co sites~\cite{inomata.prb1981}.
Moreover, the trend of MAE in Ce$_2T_2$Co$_{15}$, as shown in
\rfig{fig:mae_t2_all}, is rather generic. We also found the similar
trend in Y${_2}T_2$Co$_{15}$ and La${_2}T{_2}$Co$_{15}$, MAE increases
with $T$=Mn, or late $3d$ elements. Calculations using FP-LMTO method
also shows similar trends of MAE.

\begin{figure}[ht]
\begin{tabular}{c}
  \includegraphics[width=.42\textwidth,clip,angle=0]{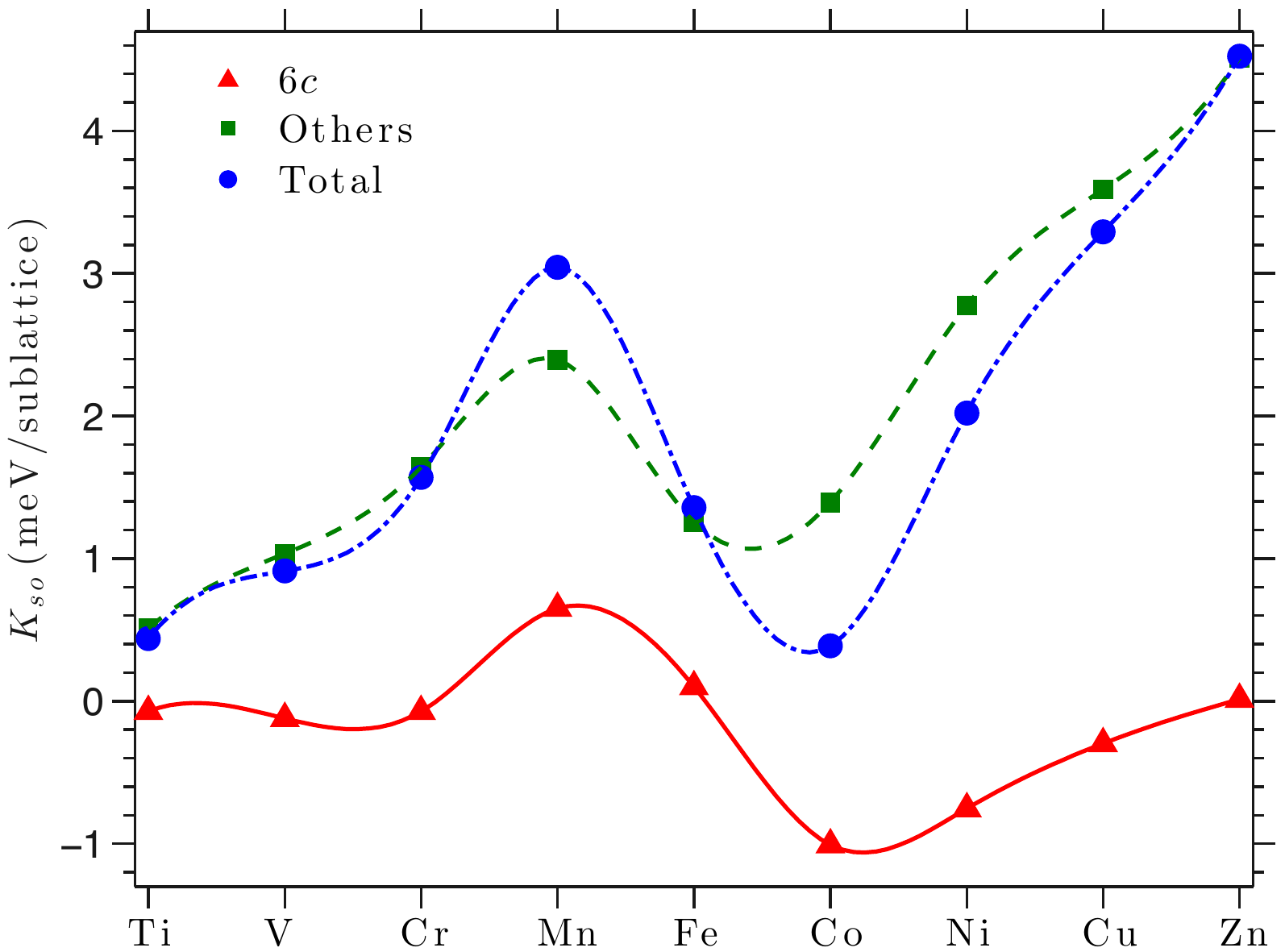}
  \\
\end{tabular}%
\caption{ Anisotropy of the scaled on-site SOC energy {\kso} in
  Ce$_2T_2$Co$_{15}$ and its contributions from the dumbbell
  sublattice $T$($6c$) and the rest sublattices. }
\label{fig:kso_t2_all}
\end{figure}

The total{} \kso, its contribution from the dumbbell site, and the
other sublattices' contributions are shown in
\rfig{fig:kso_t2_all}. Total {\kso} closely follows $K$ for all doping
elements, thus validating our use of {\kso} to resolve the MAE and
understand its origin. As shown in \rfig{fig:kso_t2_all}, the Co
dumbbell sublattice in {\rceco} has a very negative contribution to
the uniaxial anisotropy{}
\kso$(6c)$=\SI{1}{\mevfu}~(\SI{0.5}{\mevat}). Replacing Co with other
$3d$ elements decreases or eliminates this negative contribution, or
even make it positive, as with $T$=Mn. For the dumbbell site
contributions, only four elements with large magnetic moments (all
ferromagneticlly couple to Co sublattice), Mn, Fe, Co, and Ni, have
non-trivial contributions. Atoms on both ends of the $3d$ elements
have negligible contributions to the uniaxial anisotropy as
expected. Although Cu and Zn have the largest SOC constants among
$3d$, they are nearly non-magnetic, hence, they barely contribute to
the MAE itself~\cite{ke.prb2015}. The light elements Ti, V, and Cr
have small spin moments between 0.36 and 0.55{\mub} (antiparallel to
the Co sublattice) and smaller SOC constants, together resulting in a
small{} \kso($T$).

Although the dumbbell site contribution dominates the MAE enhancement
for $T$=Fe and Mn, it is obvious that the variation of MAE is a
collective effect, especially for $T$=Cu, or Zn. While the
\SI{-1}{\mevfu} negative contribution from the dumbbell sublattice is
eliminated with $T$=Cu and Zn, the contributions from the rest
sublattices increase by about 2 and \SI{3}{\mevfu},
respectively. Similarly, for the doping of non-magnetic Al atoms, the
calculated MAE in Ce$_2$Al$_2$Co$_{15}$ has a large value of
$K=\SI{3.8}{\mevfu}$. Experimentally, Al atoms had been found to
prefer to occupy the dumbbell site and also increase the uniaxial
anisotropy~\cite{inomata.prb1981,shen.jap1999}. MAE often depends on
subtle features of the bandstructure near the Fermi level; therefore,
the collective effect of MAE variation should be expected for a
metallic system~\cite{ke.prb2016a}. The modification of one site, such
as doping, unavoidably affects the electronic configuration of other
sites and their contribution to MAE.

\subsection{Origin of MAE in Ce$_2T_2$Co$_{15}$ with $T$=Fe and Mn}

\begin{figure}[ht]
\begin{tabular}{c}
\includegraphics[width=.42\textwidth,clip,angle=0]{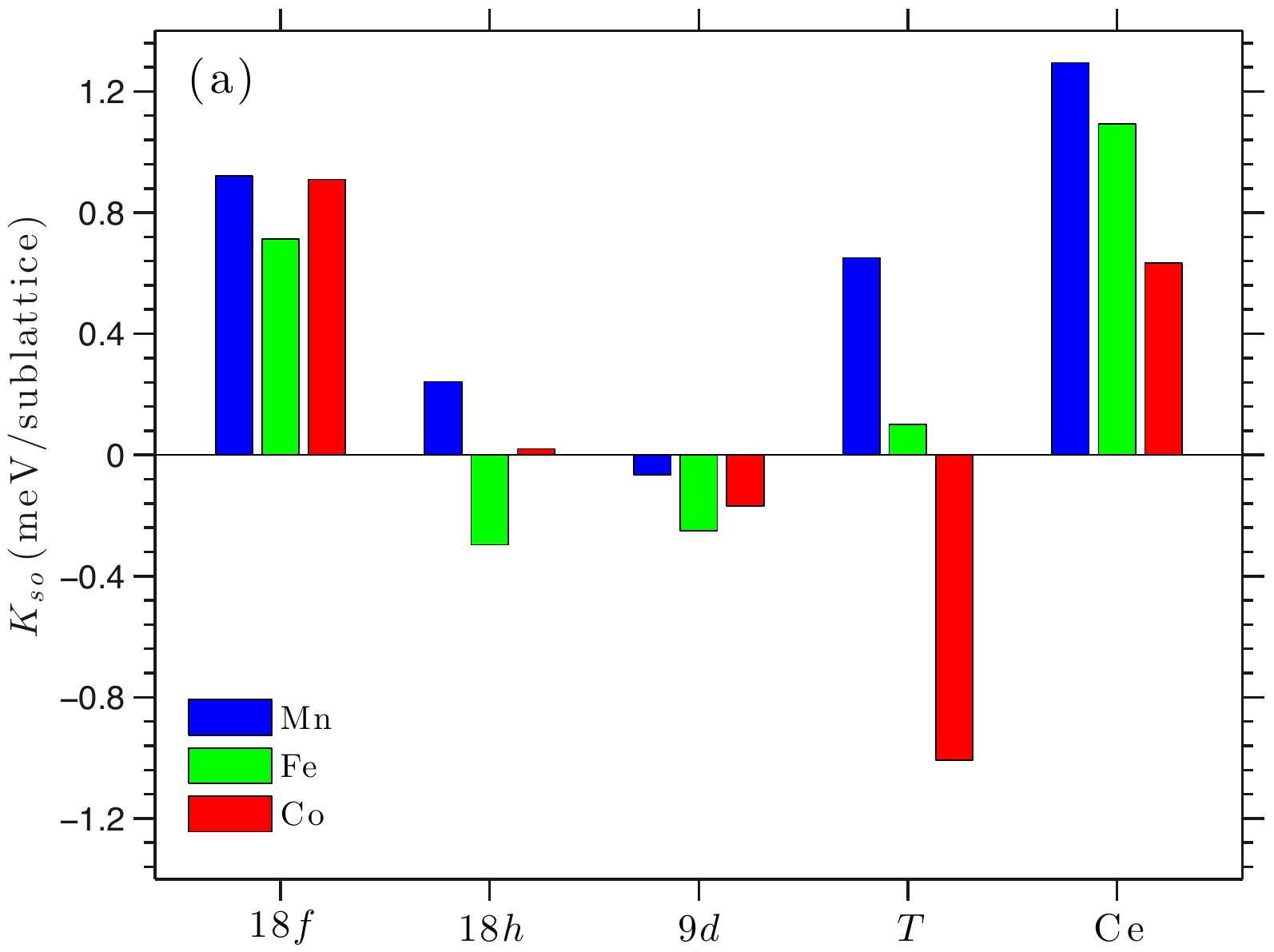} \\ 
\includegraphics[width=.42\textwidth,clip,angle=0]{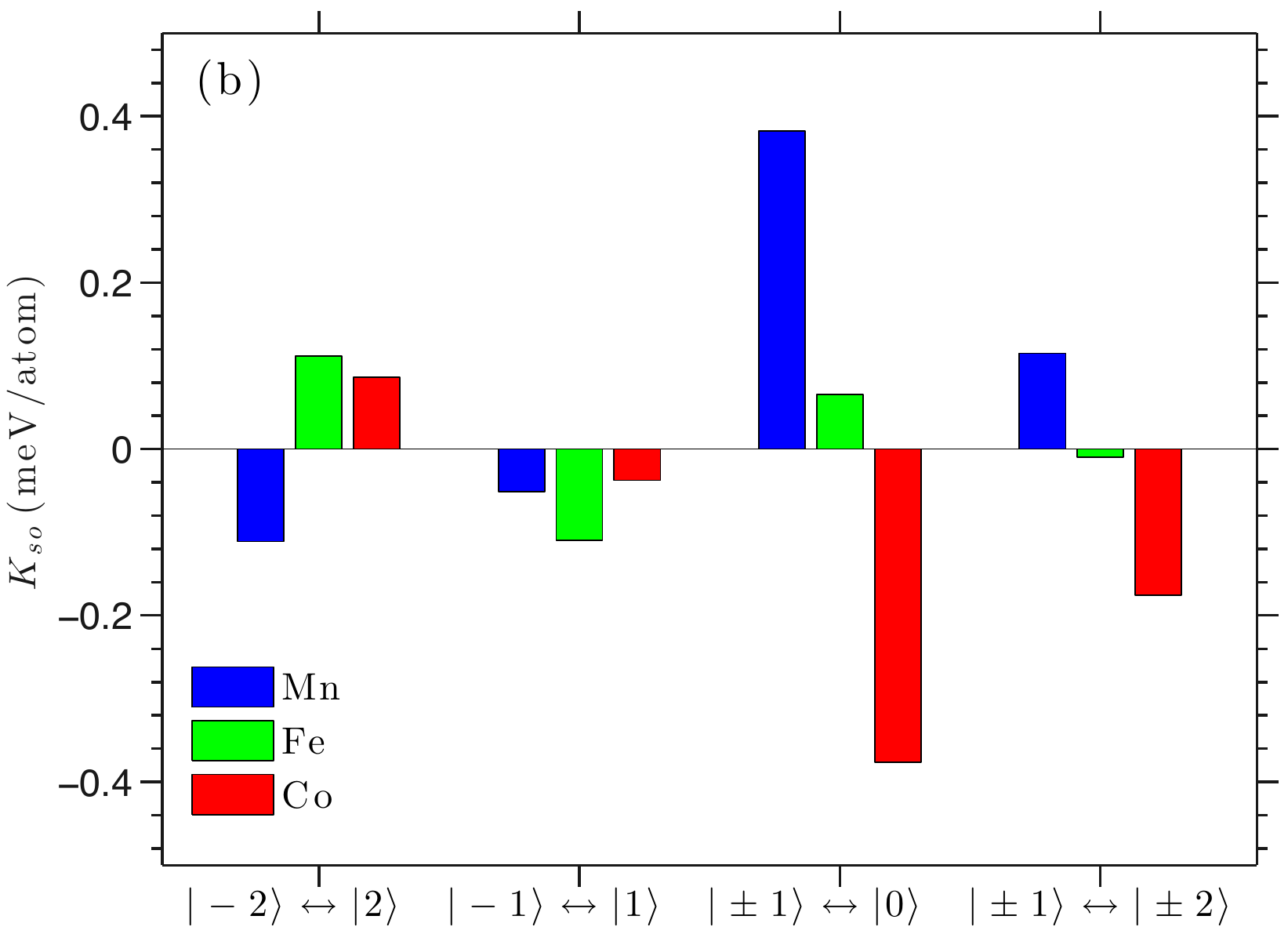}\\
\end{tabular}%
\caption{ (a) Site-resolved anisotropy of the on-site SOC energy
  {\kso} and (b) orbital-resolved {\kso($6c$)} in
  Ce$_2$Co$_{17-x}T_{x}$ with $T$=Co, Fe, and Mn. }
\label{fig:kso_site_6c}
\end{figure}

We found that all dopings except Fe and Mn decrease the magnetization,
which is consistent with the experiments by Fujji
{\etal}~\cite{fujii.jap1982}, and Schaller
{\etal}~\cite{schaller.jap1972}. Ce$_2$Fe$_{2}$Co$_{15}$ and
Ce$_2$Mn$_{2}$Co$_{15}$ have slightly larger magnetization than
{\ceco} by $5\%$ and $8\%$, respectively. It is worth noting that
experimental result on Mn doping is rather inconclusive. A slight
decrease of magnetization with Mn doping has also been
reported~\cite{sun.jpcm2000}.

Sublattice-resolved {\kso} in Ce$_2T_2$Co$_{15}$ for $T$=Co, Fe, and
Mn are shown in \rfig{fig:kso_site_6c}(a). The dominant enhancement of
MAE are from the dumbbell site, although contributions from other
sublattices also vary with $T$. To understand this enhancement of
{\kso} from the dumbbell sites, we further resolved {\kso} into
contributions from allowed transitions between all pairs of
subbands. The dumbbell sites have $3m$ symmetry. Without considering
SOC, five $d$ orbitals on $T$ sites split into three groups: $d_{z^2}$
state, degenerate ($d_{yz}$, $d_{xz}$) states, and degenerate
($d_{xy}$, $d_{x^{2}-y^{2}}$) states. Equivalently, they can be
labeled as $m$=$0$, $m$=$\pm1$, and $m$=$\pm2$ using cubic
harmonics. {\kso($T$)} can be written as~\cite{ke.prb2015}

\begin{equation}
K_\text{so}(T)=\frac{\xi^2}{4} \left(
4\mathbf{\chi}_{22}^{\epsilon}+\mathbf{\chi}_{11}^{\epsilon
}-3\mathbf{\chi}_{01}^{\epsilon}-2\mathbf{\chi}_{12}^{\epsilon}%
\right),
\label{eq:k3l}%
\end{equation}
where $\xi$ is the SOC constant and $\chi_{mm'}^{\epsilon}$ is the
difference between the spin-parallel and spin-flip components of
orbital pair susceptibility. It can be written as

\begin{equation}
\chi_{mm'}^{\epsilon}=\chi_{mm'}^{\uparrow\uparrow}+\chi_{mm'}^{\downarrow\downarrow}-\chi_{mm'}^{\uparrow\downarrow}-\chi_{mm'}^{\downarrow\uparrow}.
\label{eqn_chiep}
\end{equation}

Contributions to {\kso($T$)} resolved into transitions between pairs
of subbands are shown in \rfig{fig:kso_site_6c}(b). The four
groups of transitions correspond to the four terms in
\req{eq:k3l}. The dominant effect is from $|0\rangle \leftrightarrow
|\pm1\rangle$, namely the transitions between $d_{z^2}$ and
$(d_{yz}|d_{xz})$ orbitals. This contribution is negative for $T$=Co,
nearly disappears for $T$=Fe, and even becomes positive and large for
$T$=Mn.

The interesting dependence of $|0\rangle \leftrightarrow |\pm1\rangle$
contribution on $T$ can be understood by investigating how the
electronic structure changes with different $T$ elements. The sign of
the MAE contribution from transitions between a pair of subbands
$|m,\sigma\rangle$ and $|m',\sigma'\rangle$ is determined by the spin
and orbital character of the involved
orbitals~\cite{daalderop.prb1994,ke.prb2015}. Inter-$|m|$ transitions
$|0\rangle \leftrightarrow |\pm1\rangle$ promote easy-plane anisotropy
within the same spin channel and easy-axis anisotropy when between
different spin channels.

\begin{figure}[ht]
\begin{tabular}{c}
\includegraphics[width=.48\textwidth,clip,angle=0]{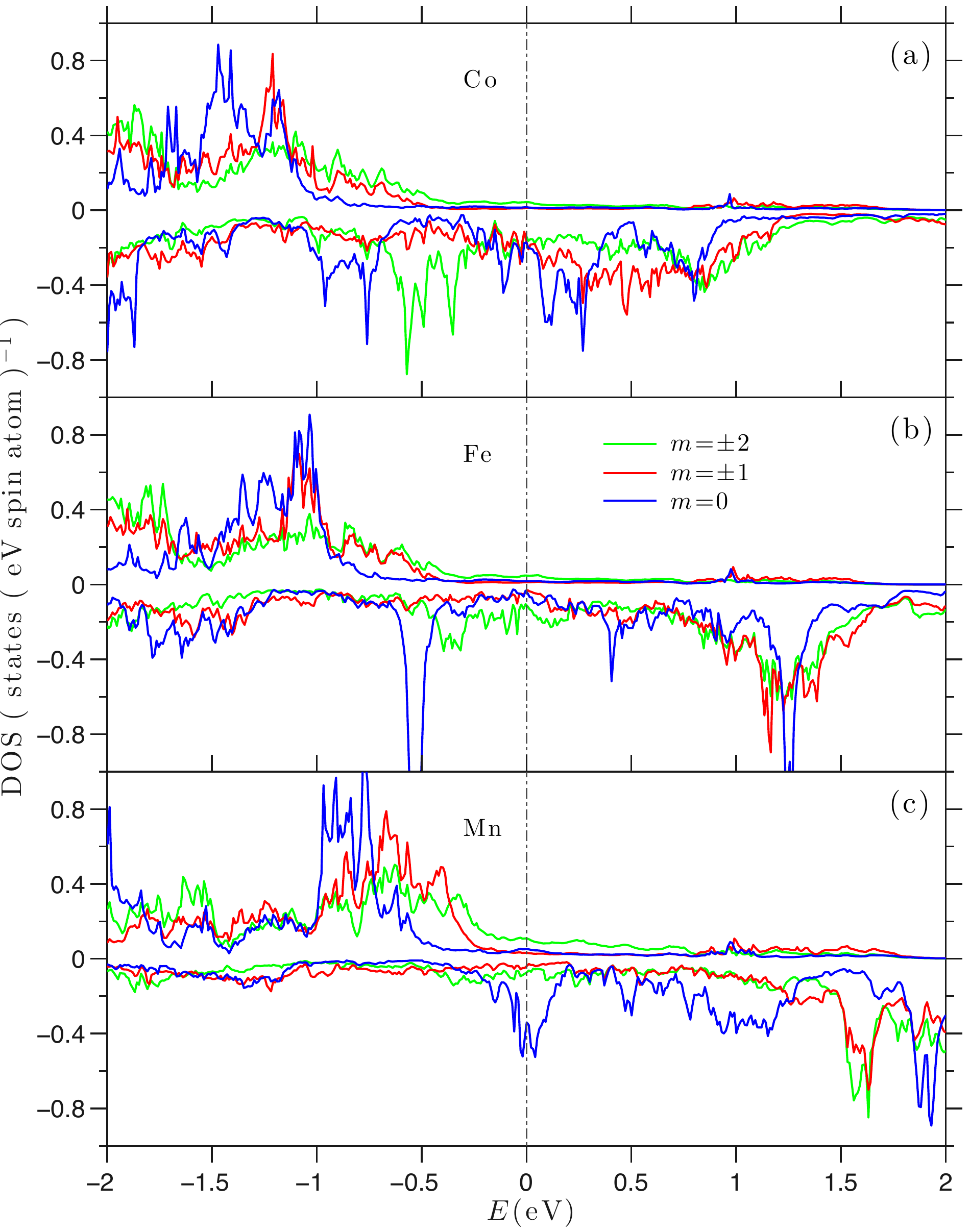}%
\end{tabular}%
\caption{ The scalar-relativistic partial density of states projected
  on the $3d$ states of $T$ sites in $R$-Ce$_2T_2$Co$_{15}$ with
  $T$=Co, Fe, and Mn. $T$ atoms occupy the dumbbell ($6c$) sites.}
\label{fig:pdos}
\end{figure}

The scalar-relativistic partial densities of states (PDOS) projected
on the dumbbell site are shown in \rfig{fig:pdos}.  For $T$=Co, the
majority spin channel is nearly fully occupied and has very small DOS
around the Fermi level, while the minority spin channel has a larger
DOS. The transitions between $d_{z^2}$ and $(d_{yz},d_{xz})$ states
across the Fermi level and within the minority spin channel, namely
$|0,\downarrow\rangle \leftrightarrow |\pm1,\downarrow\rangle$,
promote the easy-plan anisotropy. For $T$=Fe, the PDOS of $d_{z^2}$
and $(d_{yz},d_{xz})$ are rather small near the Fermi level in both
spin channels and the net contribution from $|0\rangle \leftrightarrow
|\pm1\rangle$ becomes negligible. For $T$=Mn, the Fermi level
intersects a large peak of the $d_{z^2}$ state at the Fermi level in
the minority spin channel. The spin-flip transitions
$|0,\downarrow\rangle \leftrightarrow |\pm1,\uparrow\rangle$ give rise
to a large positive contribution to uniaxial anisotropy.

\subsection{Zr, Ti, and Hf doping in Ce$_2$Co$_{17}$}

The failure to reproduce high anisotropy introduced by other dopants,
such as Zr, Ti, and V, is likely due to our oversimplified assumption
that a pair of $T$ atoms always replaces a pair of Co dumbbell atoms.
Unlike Fe and Mn, the site occupancy preference for those dopants is
not well understood~\cite{larson.prb2004}. Considering Zr doping most
effectively enhanced $H_\text{A}$ in experiments, here we focus on Zr
doping.

Both volume and chemical effects likely play important roles in
substitution site preference. To have a better understanding of the Zr
site preference, we calculated the formation energy of
Ce$_2$ZrCo$_{16}$ with the Zr atom occupying one of the four
non-equivalent Co sites and found that Zr also prefers to occupy the
dumbbell sites -- likely due to the relatively large volume around the
dumbbell sites.  The formation energies are higher by 39, 58, and
\SI{81}{\mevat} when Zr occupies the $18f$, $18h$, or $9d$ sites,
respectively.  Considering Zr atoms are relatively large, we
investigated another scenario by replacing the pair of Co dumbbell
atoms with a single Zr atom, as suggested by Larson and
Mazin~\cite{larson.prb2004}. Indeed, this latter configuration of
Ce$_2$ZrCo$_{15}$ has the lowest formation energy, which is
\SI{3}{\mevat} lower than that of Ce$_2$Zr$_2$Co$_{15}$ and
\SI{1}{\mevat} lower than Ce$_2$Co$_{16}$Zr (with Zr replacing one of
the two dumbbell Co atoms in \ceco).  That is, with Zr additions the
{\cecof} structure is preferred over the Ce$_{2}$Co$_{17}$-based
structure.  The resulting Ce$_2$ZrCo$_{15}$ has a 1-5 structure
(Ce$_{0.67}$Zr$_{0.33}$)Co$_{5}$, with one-third of the Ce in the
CeCo$_5$ structure, shown in \rfig{fig:xtal}(a), replaced by Zr atoms.
Hence, the formation energy calculation indicate that the realized
structure is likely a mix of 2-17 and 1-5 structures. Interestingly,
this may be related to experimental observations that successful 2-17
magnets usually have one common microstructure, i.e., separated cells
of 2-17 phase surrounded by a thin shell of a 1-5 boundary phase, and
Zr, Hf, or Ti additions promote the formation of such
structure~\cite{hmm.v04c2}.

The calculated anisotropy in Ce$_2$ZrCo$_{15}$, or equivalently
(Ce$_{0.67}$Zr$_{0.33}$)Co$_{5}$, is about 4{\mjmc} and much larger
than that of Ce$_2$Zr$_2$Co$_{15}$. Analysis of {\kso} reveals that
not only is the negative contribution from the previous dumbbell sites
eliminated, but more importantly, the pure Co plane becomes very
uniaxial. For $T$=V and Ti, the calculated MAE in this configuration
is also much larger than that of Ce$_2T_2$Co$_{15}$, as shown in
\rfig{fig:mae_t2_all}.  Similarly, a large MAE of \SI{2.41}{\mevfu}
was obtained for (Ce$_{0.67}$Hf$_{0.33}$)Co$_{5}$.

\section{Conclusion}
Using density functional theory, we investigated the origin of
anisotropy in doped {\ceco}. We confirmed that the dumbbell sites have
a very negative contribution to the MAE in {\ceco} with a value about
\SI{0.5}{\mevat}. The enhancement of MAE due to Fe and Mn doping
agrees well with experiments, which can be explained by the
preferential substitution effect because the enhancement is dominated
by dumbbell sites. The transitions between the $d_{z^2}$ and
($d_{yz}|d_{xz}$) subbands on dumbbell sites are responsible for the
MAE variation, and these transitions can be explained by the PDOS
around the Fermi level, which in turn depends on the element $T$
occupying on the dumbbell site.  For Zr doping, the calculated
formation energy suggests that the real structure is likely a mix of
2-17 and 1-5 structures, and the resulted 1-5 structure has a large
anisotropy, which may explain the large MAE enhancement observed in
experiments. The variation of MAE due to doping is generally a
collective effect. Doping on dumbbell sites may significantly change
the contributions from other sublattices and then the overall
anisotropy. It is worth investigating other non-magnetic elements with
a strong dumbbell site substitution preference because it may increase
the total anisotropy in this system by increasing the contributions
from other sublattices.

\section{Acknowledgments}
We thank B. Harmon, T. Hoffmann, M. K. Kashyap, R. W. McCallum, and
V. Antropov for helpful discussions. Work at Ames Laboratory was
supported by the U.S. Department of Energy, ARPA-E (REACT Grant
No. 0472-1526).  The relative stability and formation energy
investigation were supported by Office of Energy Efficiency and
Renewable Energy (EERE) under its Vehicle Technologies Program. Ames
Laboratory is operated for the U.S. Department of Energy by Iowa State
University under Contract No. DE-AC02-07CH11358.

\bibliography{aaa}

\bigskip

\end{document}